\documentclass[acmsmall,screen]{acmart}
\usepackage{multirow}
\usepackage{multicol}
\usepackage{array}    
\usepackage{framed}
\usepackage{hyperref}
\usepackage[capitalise,noabbrev]{cleveref}
\usepackage{listings}
\usepackage{xspace}
\usepackage{csquotes}
\usepackage{dblfloatfix}
\usepackage{placeins}
\usepackage{soul}
\usepackage{microtype}
\usepackage{subcaption}
\usepackage{etoolbox}
\usepackage{colortbl}
\usepackage{makecell}
\usepackage{pifont}

\definecolor{color-a}{RGB}{244, 241, 222}
\definecolor{color-b}{RGB}{129, 178, 154}
\definecolor{color-c}{RGB}{61, 64, 91}
\definecolor{color-d}{RGB}{242, 204, 143}
\definecolor{color-e}{RGB}{224, 122, 95}
\definecolor{color-f}{RGB}{201, 228, 202}
\definecolor{color-g}{RGB}{254, 217, 183}
\definecolor{gray}{gray}{0.9}
\definecolor{light-gray}{gray}{0.4}

\newcommand{\dataset}{\textsc{CleanVul}\xspace}
\newcommand{\tool}{\textsc{VulSifter}\xspace}

\newcommand{\cmark}{\ding{51}}%
\newcommand{\xmark}{\ding{55}}%
\newcommand{\mycmark}{\color{color-b} \cmark \color{black}}

\newcommand{\myxmark}{\color{color-e} \xmark \color{black}}

\sethlcolor{cyan}

\renewcommand{\paragraph}[1]{\vspace{6pt}\noindent{\bf #1}\hspace{8pt}}
\newcommand{\mytextcolor}[2]{{\sethlcolor{#1}\hl{#2}}}
\newcommand{\mycodecolor}[2]{\mytextcolor{#1}{\textls{#2}}}

\newcommand\realnumberstyle[1]{}
\makeatletter
\newcommand{\linecolor}[3]{
    {\realnumberstyle{#3}}
    \begingroup
    \lst@basicstyle
    \ifnum\value{lstnumber}=#1
        \color{#2}
    \else
        \color{white}
    \fi
    \rlap{\hspace*{\lst@numbersep}
    \color@block{\linewidth}{\ht\strutbox}{\dp\strutbox}
    }
    \endgroup
}

\makeatletter
\newcommand{\storelinecolor}[2]{%
    \expandafter\gdef\csname linecolor@#1\endcsname{#2}%
}

\newcommand{\clearlinecolors}{%
    \@for\next:=1,2,3,4,5,6,7,8,9\do{%
        \expandafter\global\expandafter\let\csname linecolor@\next\endcsname\undefined
    }%
}

\newcommand{\multilinecolor}[1]{%
    {\realnumberstyle{#1}}%
    \begingroup
    \lst@basicstyle
    \ifcsname linecolor@\arabic{lstnumber}\endcsname
        \color{\csname linecolor@\arabic{lstnumber}\endcsname}%
    \else
        \color{white}%
    \fi
    \rlap{\hspace*{\lst@numbersep}%
    \color@block{\linewidth}{\ht\strutbox}{\dp\strutbox}%
    }%
    \endgroup
}

\newcommand{\setlinecolors}[1]{%
    \clearlinecolors
    #1%
}
\makeatother

\newcommand{\linecolorrg}[3]{
    {\realnumberstyle{#3}}
    \begingroup
    \lst@basicstyle
    \ifnum\value{lstnumber}=#1
        \color{#2}
    \else
        \ifnum\value{lstnumber}>#1
            \ifnum\value{lstnumber}<\numexpr#1+\@firstofone#2\relax
                \color{\@secondoftwo#2}
            \else
                \color{white}
            \fi
        \else
            \color{white}
        \fi
    \fi
    \rlap{\hspace*{\lst@numbersep}
    \color@block{\linewidth}{\ht\strutbox}{\dp\strutbox}
    }
    \endgroup
}

\newcommand{\linecolorr}[4]{
    {\realnumberstyle{#4}}
    \begingroup
    \lst@basicstyle
    \ifnum\value{lstnumber}>=#1
        \ifnum\value{lstnumber}<=#2
            \color{#3}
        \else
            \color{red}
        \fi
    \else
        \color{white}
    \fi
    \rlap{\hspace*{\lst@numbersep}
    \color@block{\linewidth}{\ht\strutbox}{\dp\strutbox}
    }
    \endgroup
}

\newcommand{\linecolorrange}[4]{
    {\realnumberstyle{#4}}
    \begingroup
    \lst@basicstyle
    \ifnum\value{lstnumber}>=#1
        \ifnum\value{lstnumber}<=#2
            \color{#3}
        \else
            \color{white}
        \fi
    \else
        \color{white}
    \fi
    \rlap{\hspace*{\lst@numbersep}
    \color@block{\linewidth}{\ht\strutbox}{\dp\strutbox}
    }
    \endgroup
}

\newcommand{\colorlines}[1]{
    \renewcommand{\lst@DefEC}{%
        \lst@CCECUse \lst@ProcessLetter
        \global\let\lst@thestyle\relax
        \edef\tempa{#1}\edef\tempb{\thelstnumber}%
        \ifstrequal{\tempa}{\tempb}%
           {\global\def\lst@thestyle{\color{red}}}{}%
    }
    \expandafter\lst@AddToHook\expandafter{OutputOther}{\lst@thestyle}
}

\lstdefinelanguage{Diff}{
  language=Python,
  sensitive=true,
  morecomment=[f][\color{myred}]-,
  morecomment=[f][\color{mygreen}]+,
}

\AtBeginDocument{%
  }

\setcopyright{acmlicensed}
\copyrightyear{2018}
\acmYear{2018}
\acmDOI{XXXXXXX.XXXXXXX}

\acmJournal{JACM}
\acmVolume{37}
\acmNumber{4}
\acmArticle{111}
\acmMonth{8}

\begin{document}

\title{\dataset: Toward High-Quality Function-Level Vulnerability Datasets via LLM-Based Noise Reduction}

\author{Yikun Li}
\email{yikunli@smu.edu.sg}
\orcid{https://orcid.org/0000-0002-1566-725X}
\affiliation{%
  \institution{Singapore Management University}
  \city{Singapore}
  \country{Singapore}
}

\author{Ting Zhang}
\affiliation{%
  \institution{Singapore Management University}
  \city{Singapore}
  \country{Singapore}
}

\author{Ratnadira Widyasari}
\affiliation{%
  \institution{Singapore Management University}
  \city{Singapore}
  \country{Singapore}
}

\author{Yan Naing Tun}
\affiliation{%
  \institution{Singapore Management University}
  \city{Singapore}
  \country{Singapore}
}

\author{Huu Hung Nguyen}
\affiliation{%
  \institution{Singapore Management University}
  \city{Singapore}
  \country{Singapore}
}

\author{Tan Bui}
\affiliation{%
  \institution{Singapore Management University}
  \city{Singapore}
  \country{Singapore}
}

\author{Ivana Clairine Irsan}
\affiliation{%
  \institution{Singapore Management University}
  \city{Singapore}
  \country{Singapore}
}

\author{Yiran Cheng}
\affiliation{%
  \institution{Singapore Management University}
  \city{Singapore}
  \country{Singapore}
}

\author{Xiang Lan}
\affiliation{%
  \institution{North Carolina State University}
  \city{Raleigh, North Carolina}
  \country{United States}
}

\author{Han Wei Ang}
\affiliation{%
  \institution{GovTech}
  \city{Singapore}
  \country{Singapore}
}

\author{Frank Liauw}
\affiliation{%
  \institution{GovTech}
  \city{Singapore}
  \country{Singapore}
}

\author{Martin Weyssow}
\affiliation{%
  \institution{Singapore Management University}
  \city{Singapore}
  \country{Singapore}
}

\author{Hong Jin Kang}
\affiliation{%
  \institution{Singapore Management University}
  \city{Singapore}
  \country{Singapore}
}

\author{Eng Lieh Ouh}
\email{elouh@smu.edu.sg}
\affiliation{%
  \institution{Singapore Management University}
  \city{Singapore}
  \country{Singapore}
}

\author{Lwin Khin Shar}
\email{lkshar@smu.edu.sg}
\affiliation{%
  \institution{Singapore Management University}
  \city{Singapore}
  \country{Singapore}
}

\author{David Lo}
\email{davidlo@smu.edu.sg}
\orcid{https://orcid.org/0000-0002-4367-7201}
\affiliation{%
  \institution{Singapore Management University}
  \city{Singapore}
  \country{Singapore}
}

\renewcommand{\shortauthors}{Li et al.}

\begin{abstract}
In the dynamic field of cybersecurity, the accurate identification and mitigation of software vulnerabilities are paramount for maintaining system integrity. Vulnerability datasets, often derived from the National Vulnerability Database (NVD) or directly from GitHub, are essential for training machine learning models to detect and address these security flaws. However, these datasets frequently suffer from significant noise, typically 40\% to 75\% \cite{ding2024vulnerability,chen2023diversevul}, due primarily to the automatic and indiscriminate labeling of all modifications in vulnerability-fixing commits (VFCs) as vulnerability-related. This misclassification occurs because not all changes in a commit aimed at fixing vulnerabilities pertain to security threats; many are routine updates like bug fixes, test improvements, or unrelated code refactoring.

To address these challenges, this paper introduces the \textbf{first methodology} that leverages the Large Language Model (LLM) with a heuristic enhancement to automatically identify vulnerability-fixing changes from VFCs, achieving an F1-score of 0.82. \tool was applied to a large-scale study, where we conducted a comprehensive crawl of 127,063 repositories on GitHub, resulting in the acquisition of 5,352,105 commits. The LLM with heuristic enhancement approach involves utilizing an LLM to comprehend code semantics and contextual information, while applying heuristics to filter out unrelated changes. We developed \dataset, a high-quality dataset comprising 8,198 functions using our LLM heuristic enhancement approach, demonstrating \emph{Correctness} (90.6\%) comparable to established datasets such as SVEN (94.0\%) \cite{he2023large} and PrimeVul (86.0\%) \cite{ding2024vulnerability}. In this context, \emph{Correctness} is defined as the percentage of genuine vulnerable functions in the vulnerability dataset, which helps assess the effectiveness of \tool in identifying and filtering vulnerability-fixing changes. To evaluate the effectiveness of our \dataset dataset, we conducted experiments focusing on fine-tuning various LLMs on the \dataset dataset and other high-quality datasets. Our evaluation was conducted on multiple popular programming languages, including Java, Python, JavaScript, C\#, C, and C++. Evaluation results reveal that LLMs fine-tuned on \dataset not only exhibit enhanced accuracy but also superior generalization capabilities compared to those trained on uncleaned datasets. Specifically, models trained on \dataset and tested on PrimeVul achieve accuracy higher than those trained and tested exclusively on PrimeVul, validating the effectiveness of our methodology.
\end{abstract}



\received{20 February 2007}
\received[revised]{12 March 2009}
\received[accepted]{5 June 2009}

\maketitle

\section{Introduction}

In the rapidly evolving field of cybersecurity, accurately detecting and mitigating software vulnerabilities is crucial for safeguarding digital assets and infrastructure. These vulnerabilities pose significant security risks, underscoring the need for robust tools to identify and rectify them. Vulnerability datasets serve as crucial resources for training machine learning (ML) models to identify and address these security flaws \cite{chakraborty2021deep, chen2023diversevul, steenhoek2023empirical, fu2022linevul}. These datasets are typically derived from known vulnerabilities cataloged in databases like the National Vulnerability Database (NVD) or mined from software repositories such as GitHub. By leveraging these datasets, ML models can learn patterns and indicators of security vulnerabilities, thereby enhancing automated vulnerability detection systems.

However, existing vulnerability datasets \cite{lu2021codexglue, chen2023diversevul, fan2020ac, bhandari2021cvefixes, nikitopoulos2021crossvul} often contain significant noise, ranging from 40\% to 75\%, which substantially reduces the training effectiveness of ML models \cite{chen2023diversevul,ding2024vulnerability}. This issue arises because many datasets automatically label all modifications in vulnerability-fixing commits (VFCs) as related to vulnerabilities, failing to recognize that not all changes pertain to security threats. 
Frequently, a commit intended to address a specific vulnerability might also include unrelated code adjustments. This leads to the erroneous classification of such benign modifications as security threats. 
A detailed example illustrating this challenge is provided in \cref{sec:motivation}.

To address these shortcomings, several attempts have been made to enhance dataset accuracy by correlating function names from commit logs with the descriptions found in the NVD \cite{ding2024vulnerability}. Specifically, a function is labeled as vulnerable if it is explicitly mentioned in the NVD description or if it is the only function changed in the file mentioned by the NVD. However, this method falls short when applied to VFCs that lack corresponding NVD entries, a common scenario for most of VFCs identified on GitHub. Consequently, there is a critical need to develop robust, automated techniques capable of discerning genuine vulnerability-fixing changes across the entirety of VFCs, regardless of their NVD linkage.

\paragraph{Our Solution}
This paper introduces the first methodology to address the gap in accurately identifying vulnerability-related changes within VFCs. 
We first examine the reasons why many changes cataloged as vulnerability fixes do not pertain to actual security vulnerabilities. 
Our empirical study reveals that approximately 80\% of non-vulnerability changes consist of test-related modifications and general bug fixes, with additional changes spanning support updates, code refactoring, and documentation updates.
Building on this analysis, we developed \textbf{\tool}, a novel approach that combines LLM with heuristic filtering to automatically analyze VFCs and eliminate noise, resulting in a refined vulnerability dataset called \textbf{\dataset}.
By applying \tool, the \emph{Correctness} of genuine vulnerability fixes in the dataset improved from 28.7\% to 90.6\%. \emph{Correctness} is defined as the percentage of genuine vulnerable functions in the vulnerability dataset, and it helps assess the effectiveness of \tool in identifying and filtering vulnerability-fixing changes. 
\dataset comprises 8,198 functions, both vulnerable and benign, and achieves a level of \emph{Correctness} (90.6\%) comparable to established datasets such as SVEN~\cite{he2023large} and PrimeVul~\cite{ding2024vulnerability}. 
For comparison, the manually curated SVEN dataset contains 803 functions with 94.0\% \emph{Correctness}, while PrimeVul, which identifies vulnerabilities by matching function names with NVD descriptions, contains 6,968 NVD-sourced functions with 86.0\% \emph{Correctness}.
As the first automated approach to improve dataset \emph{Correctness} without requiring NVD linkage or other constraints, \dataset represents a significant advancement in function-level vulnerability detection. 
It complements PrimeVul's NVD-based approach by automatically filtering noise from GitHub VFCs to obtain high-quality vulnerability datasets.

\paragraph{Effectiveness of \dataset} 
We evaluated our dataset \dataset by comparing it with two established high-quality vulnerability detection datasets: SVEN and PrimeVul. 
Since our dataset consists of balanced vulnerability pairs (vulnerable/benign), we use accuracy as our primary evaluation metric rather than F1-score, as accuracy better reflects model performance when classes are balanced - a random baseline would achieve 50\% accuracy. 
Through experiments with various LLMs, we found that models fine-tuned on \dataset showed strong generalization capabilities. When testing on PrimeVul, our \dataset-trained models achieved 58.09\% accuracy (using CodeBERT), exceeding PrimeVul's own intra-dataset performance of 56.61\%. When testing on SVEN, \dataset-trained models reached 64.87\% accuracy, outperforming models fine-tuned on PrimeVul which only achieved 55.75\% accuracy on SVEN. Additionally, \dataset demonstrated robust intra-dataset performance with 68.96\% accuracy when trained and tested on itself. These results suggest that \dataset captures a more diverse and representative range of vulnerabilities compared to existing datasets, making it particularly effective for training vulnerability detection models.

\paragraph{Contributions} 
Overall, we make the following contributions:

\begin{itemize}
    \item \textbf{Characterization of Code Change Categories in VFCs:}
    We conducted a manual analysis to categorize non-vulnerability-related changes within VFC datasets. We also developed a refined taxonomy capturing the nuances of function-level changes, revealing that the predominant non-vulnerability changes were test-related (41.2\%) and bug fixes (38.2\%).

    \item \textbf{\tool~- LLM Heuristic for Identifying Vulnerability Fixes in VFCs:}
    We proposed and validated the first approach to automatically identify function-level vulnerability-fixing changes, namely \tool. We then demonstrated that LLM heuristic is particularly effective when analyzing function changes combined with commit messages and additional context, with GPT-4 achieving the highest F1-score of 0.82.

    \item \textbf{\dataset~- A Large-Scale, High-Quality Vulnerability Dataset:}
    We introduced \dataset, a new high-quality dataset derived from the application of the heuristic approach, containing 8,198 functions categorized as vulnerable and benign. We also showed that \dataset maintains \emph{Correctness} level (percentage of genuine vulnerable functions) comparable to existing high-quality datasets like SVEN and PrimeVul, enhancing the scale and reliability of data available for vulnerability detection research.

    \item \textbf{Evaluation and Comparison of Dataset Effectiveness:}
    We assessed the performance of LLMs trained on \dataset and compared it to performances on an uncleaned dataset and other established datasets like SVEN and PrimeVul. We then confirmed that training on \dataset improves model accuracy in comparison to the uncleaned dataset with a large margin, and highlighted the superior generalization capabilities of models trained on \dataset when tested on external datasets.
\end{itemize}

In the spirit of open science, we make our source code and dataset publicly available\footnote{\url{https://github.com/yikun-li/CleanVul}}.

\paragraph{Paper Structure} 
The remainder of the paper is organized as follows. \cref{sec:motivation} presents motivating examples for our work. 
\cref{sec:approach} describes the VulSifter approach, while \cref{sec:curation} explains the \dataset dataset curation process. 
\cref{sec:eva_settings,sec:eva} present our evaluation settings and experiments respectively. 
\cref{sec:analysis} provides additional analyses, including sensitivity analysis and ablation study, followed by threats to validity in \cref{sec:threats}. 
\cref{sec:related_work} covers related work, and \cref{sec:conclusion} concludes our paper.

\section{Motivating Example}
\label{sec:motivation}

Consider the following real-world example\footnote{\url{https://github.com/thingsboard/thingsboard-edge/commit/1a3ee8512d58625940b25e46bc6488a3539fdc5e}} from the ThingsBoard project, which illustrates the complexity of identifying vulnerability fixes within commits. The commit message is presented below, targeting XSS vulnerabilities. This commit exemplifies what researchers call \emph{tangled commits} \cite{kirinuki2014hey} - commits that address multiple concerns or include changes serving different purposes simultaneously. While the primary purpose is introducing security measures such as \emph{NoXss} validation to prevent XSS attacks, the commit also encompasses unrelated modifications including enhancing the codebase by removing unnecessary imports and updating license documentation, as well as improving test coverage to confirm the efficacy of the new validations.

\begin{lstlisting}[language=tex, numbers=left, numberstyle=\linecolor{0}{color-f}]
(*@\textbf{Fixed xss vulnerabilities in attributes and telemetry (\#8238)}@*)
* added noxss validation on kventries
* added ConstraintValidator usages for validation
* fixed licence
* added test
* removed redundant imports
\end{lstlisting}

This example underscores the challenge in analyzing commits categorized as VFCs. Assuming all modifications in such commits are responses to vulnerabilities would inaccurately label updates like documentation revisions, code cleanup, and test enhancements as security measures. This misclassification could distort the perceived security posture of the codebase and affect the effectiveness of training data for machine learning models.
Therefore, it is essential to develop automated tools capable of discerning which changes within a VFC directly address security issues and which do not. By accurately segregating these changes, we can refine datasets to include only genuine vulnerability fixes, thereby enhancing the accuracy and reliability of vulnerability detection and analysis tools.

\paragraph{Empirical Study}
Building on this motivating example, we conducted an empirical study to better understand the types of changes commonly found in VFCs.
Previous research constructing vulnerability datasets often identifies VFCs from the NVD dataset and treats all code changes within a commit as vulnerability-fixing changes \cite{ding2024vulnerability}. However, as found by PrimeVul \cite{ding2024vulnerability}, this approach is inaccurate because many commits are complex and include changes unrelated to vulnerability fixes; some commits are monolithic and encompass various types of changes. Since no prior study has deeply analyzed the changes within VFCs, we aim to characterize these changes into different categories. This characterization will: (1) help us better understand why some changes are not related to vulnerability fixes; and (2) allow us to develop heuristic-based approaches to accurately identify vulnerability-fixing changes.

We conducted a manual analysis of 136 instances where non-vulnerability-related changes occurred within the VFC datasets. Our initial classification categories were derived from previous research \cite{levin2017boosting}. We employed an open card sorting process to systematically analyze and categorize the instances. Throughout our analysis, we adapted and expanded these categories as necessary, ultimately developing a refined taxonomy that captures the nuances of function-level changes not directly related to vulnerability fixes. In the following sections, we present several reasons for these changes, accompanied by specific examples to illustrate our findings.

\paragraph{Example I: Test-Related Changes}
In some cases, developers introduce changes to the testing code when addressing vulnerabilities. These changes, while related to the vulnerability, do not directly alter the vulnerability's resolution but instead aim to verify the fix's effectiveness. For instance, in the Elasticsearch project (commit da3428), developers encountered a potential infinite loop issue when the span setting was too close to the length of the context part and the context ended in a word that tokenized to more than one token. To address this, they not only modified the core algorithm to prevent the infinite loop but also added a test case to ensure the issue was resolved:

\setlinecolors{%
    \storelinecolor{1}{color-f}%
    \storelinecolor{2}{color-f}%
    \storelinecolor{3}{color-f}%
    \storelinecolor{4}{color-f}%
    \storelinecolor{5}{color-f}%
    \storelinecolor{6}{color-f}%
    \storelinecolor{7}{color-f}%
    \storelinecolor{8}{color-f}%
    \storelinecolor{9}{color-f}%
    \storelinecolor{10}{color-f}%
    \storelinecolor{11}{color-f}%
}

\begin{lstlisting}[language=Java, numbers=left, numberstyle=\multilinecolor]
+ public void testDetectInfiniteLoop() {
+     // These settings are known to produce an infinite loop.
+     // question and context are longer than max sequence length
+     // so the input must be spanned. With a span setting of 4
+     // there is only 1 more token that can go into the context part:
      ...
+     var e = expectThrows(IllegalStateException.class, () -> tokenizer.tokenize(question, 
+         context, Tokenization.Truncate.NONE, span, 0));
+     assertThat(e.getMessage(), containsString("Tokenization cannot be satisfied with the 
+         current span setting"));
+ }
\end{lstlisting}

\paragraph{Example II: Bug Fixes}
Our analysis reveals a significant number of changes in VFCs from GitHub that were initially classified as vulnerability fixes but were actually bug fixes or feature enhancements. This discrepancy is primarily due to the use of automated keyword matching techniques \cite{bui2024javavfc} that are not always accurate and can generate false positives. Moreover, the inherent nature of commit messages themselves often limits the effectiveness of these detection methods. Many commit messages are vague and lack sufficient information to conclusively determine whether a change is related to vulnerability fixes. This ambiguity necessitates further examination, such as reviewing the actual code changes or consulting issue tracking descriptions, to accurately classify a commit. For instance, the commit message below was flagged as a VFC, but further investigation revealed it was related to gameplay features in a computer game, not a security vulnerability:

\begin{displayquote}
\textit{``Removed the attack peaceful towns validation check WHEN the peaceful feature is off''} - [SiegeWar-1fbba5]
\end{displayquote}

The presence of \emph{attack} as a keyword in VFC identification algorithms led to its incorrect classification. The code change associated with this commit further illustrates this point:

\setlinecolors{%
    \storelinecolor{2}{color-f}%
}

\begin{lstlisting}[language=Java, numbers=left, numberstyle=\multilinecolor]
- if (defendingTown.isNeutral())
+ if ((*@\mycodecolor{color-b}{SiegeWarSettings.getWarCommonPeacefulTownsEnabled() \&\&}@*) defendingTown.isNeutral())
      throw new TownyException(Translation.of("msg_war_siege_err_cannot_attack_peaceful_town"));
\end{lstlisting}

\paragraph{Example III: Support Changes}
In certain instances, developers not only fix vulnerabilities but also modify other sections of code to support these fixes, which may involve adjusting to updated dependencies or configurations. This can be seen in the following commit message, where dependencies were updated to address specific CVEs, necessitating changes in the use of affected libraries or functions:

\begin{displayquote}
\textit{``Update of direct dependencies: kubernetes java-client to 19.0.0, docker-java-bom to 3.3.4. To address CVES: CVE-2023-3635 in okio, CVE-2023-33201 in bcjava''} - [Druid-3c7dec]
\end{displayquote}

This commit indicates that the modifications were targeted at fixing vulnerabilities \emph{CVE-2023-3635} and \emph{CVE-2023-33201} through updates to two key dependencies. The updates to these packages necessitated adjustments in the code where these dependencies are used, to ensure compatibility with new versions and continued secure operation. The following code snippet illustrates the changes made to the function calls, adapting to the updated interface of the dependencies:

\setlinecolors{%
    \storelinecolor{4}{color-g}%
    \storelinecolor{5}{color-f}%
}

\begin{lstlisting}[language=Java, numbers=left, numberstyle=\multilinecolor]
  Watch.createWatch(
      realK8sClient,
      coreV1Api.listNamespacedPodCall(namespace, null, true, null, null,
-         labelSelector, null, lastKnownResourceVersion, null, 0, true, null
+         labelSelector, null, lastKnownResourceVersion, null, (*@\mycodecolor{color-b}{null}@*), 0, true, null
      ),
      new TypeReference<Watch.Response<V1Pod>>()
      {
\end{lstlisting}

\paragraph{Example IV: Code Refactoring}
During the process of addressing vulnerabilities, developers sometimes refactor the code to enhance its readability, maintainability, or extensibility. For example, in the Keycloak project (commit 15a21b), in addition to patching a vulnerability that could allow unauthorized access to user data, developers also refactored part of the code to improve its maintainability:

\setlinecolors{%
    \storelinecolor{4}{color-g}%
    \storelinecolor{5}{color-g}%
    \storelinecolor{6}{color-g}%
    \storelinecolor{8}{color-f}%
}

\begin{lstlisting}[language=Java, numbers=left, numberstyle=\multilinecolor]
  if (validRedirect.startsWith("/")) {
      validRedirect = relativeToAbsoluteURI(session, rootUrl, validRedirect);
      logger.debugv("replacing relative valid redirect with: {0}", validRedirect);
-     resolveValidRedirects.add(validRedirect);
- } else {
-     resolveValidRedirects.add(validRedirect);
  }
+ resolveValidRedirects.add(validRedirect);
\end{lstlisting}

\paragraph{Example V: Documentation Updates}
While addressing vulnerabilities, developers also take the opportunity to update documentation, enhancing the clarity and security of the code. For example, during enhancements to XML processing security to mitigate XXE attacks, javadocs were also updated:

\setlinecolors{%
    \storelinecolor{4}{color-f}%
}

\begin{lstlisting}[language=Java, numbers=left, numberstyle=\multilinecolor]
 /**
  * Configures a {@link DocumentBuilderFactory} to protect it against XML
  * External Entity attacks.
+ *
  * @param factory the factory
  * @see <a href="https://www.owasp.org/index.php/XML_External_Entity_%
\end{lstlisting}

\begin{table}[ht]
\centering
\caption{Reasons for Non-Vulnerability Changes in Identified VFCs}
\label{tb:reasons}
\begin{tabular}{@{}lcc@{}}
\toprule
\textbf{Type} & \textbf{Number (\#)} & \textbf{Percentage (\%)} \\ 
\midrule
Test-Related Changes  & 56 & 41.2 \\
Bug Fixes             & 52 & 38.2 \\
Support Changes       & 20 & 14.7 \\
Code Refactoring      &  7 &  5.1 \\
Documentation Updates &  1 &  0.7 \\
\bottomrule
\end{tabular}
\end{table}

The results are summarized in \cref{tb:reasons}. The data show that \emph{Test-Related Changes} and \emph{Bug Fixes} are the most prevalent categories of non-vulnerability changes, representing 41.2\% and 38.2\% of the cases, respectively. These two categories together account for nearly 80\% of all non-vulnerability changes in the analyzed VFCs, indicating a significant focus on functionality enhancement and reliability testing in software updates. Additionally, \emph{Support Changes} and \emph{Code Refactoring} are observed to constitute 14.7\% and 5.1\% of the changes, respectively, reflecting a lesser but noteworthy commitment to adapting existing systems and improving code quality. The smallest category, \emph{Documentation Updates}, makes up 0.7\% of the changes, suggesting minimal alterations in documentation alongside other code changes.

\section{\tool: Approach}
\label{sec:approach}

\begin{figure}[htb]
  \centering
  \includegraphics[trim={1cm 5.5cm 1cm 4.2cm}, clip, width=0.85\linewidth]{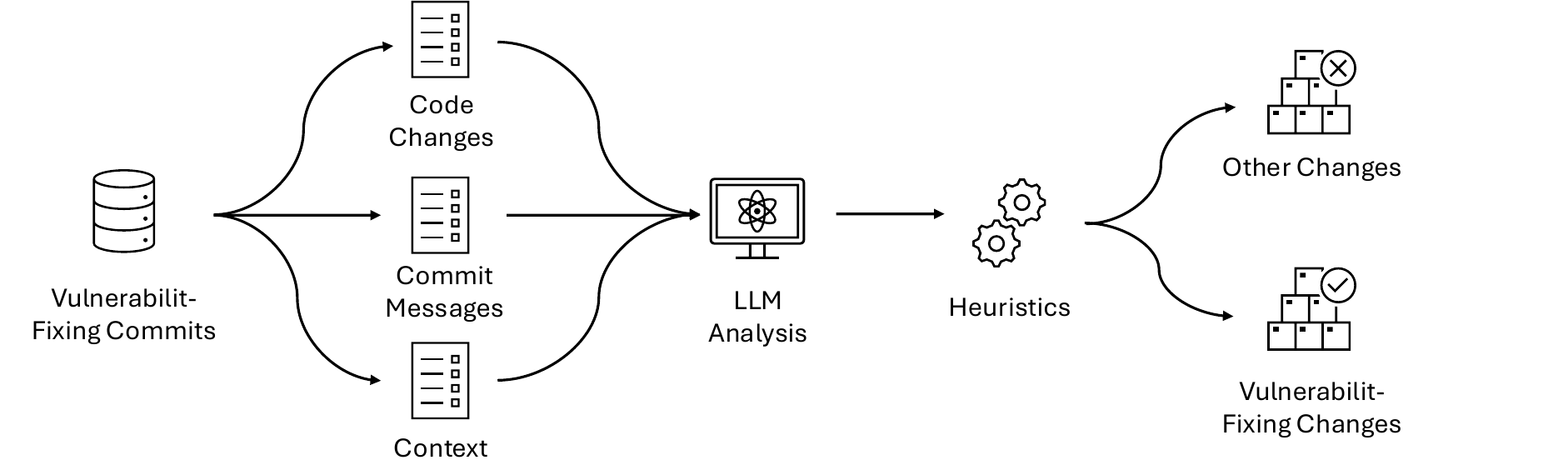}
  \caption{Overview of the Methods and Experiments Conducted in This Paper}
  \label{f:overview}
\end{figure}

The goal of this study is to analyze source code and commit messages from VFCs for the purpose of automatically identifying function-level vulnerability-fixing changes.
The overview of our approach is demonstrated in \cref{f:overview}.
The overview figure illustrates the methodological framework divided into two primary stages: \emph{LLM Analysis} and \emph{Heuristics}.

\paragraph{LLM Analysis}
We developed a prompt for LLMs that analyzes function changes along with their commit messages and context (other changed functions in the same commit). The complete prompt details are provided in \cref{fig:range_output}. 
The prompt produces a score from 0 to 4, representing the confidence in predicting vulnerability-fixing changes. To provide a clearer understanding of the scoring system, we offer the following explanations for each score:

\begin{itemize}
    \item Score 0: No vulnerability detected
    \item Score 1: Low likelihood of vulnerability
    \item Score 2: Moderate likelihood of vulnerability
    \item Score 3: High likelihood of vulnerability
    \item Score 4: Very high likelihood of vulnerability
\end{itemize}

The motivation for using the prompt outputting a range from 0 to 4 instead of common binary output stems from how this output scale allows for fine-tuned control over dataset cleanliness. In particular, an exceptionally clean dataset might exclusively comprise items scored as 4, representing the utmost confidence in vulnerability-fixing changes. In contrast, for a more comprehensive dataset, items scored 1 or higher could be retained, encompassing a wider array of potential vulnerability-fixing changes.

\begin{figure}[htb]
\begin{lstlisting}[language=Tex]
(*@\textbf{The Prompt Produces a Score from 0 to 4:}@*)
As a cybersecurity expert, analyze the provided "Original" and "Revised" code snippets from a commit along with the commit message and other functions in the same commit, where the "Original" code represents the state prior to the changes, and the "Revised" code represents the state after the changes. (*@\textbf{Evaluate the changes made in terms of vulnerability fixing on a scale of 0-4}@*). The length of the code snippet should not influence your assessment; focus on evaluating the logic line by line.

- A score of 0 indicates that the changes made from the "Original" code to the "Revised" code (*@\textbf{are not related to fixing vulnerabilities}@*).
- A score of 4 indicates that the changes made from the "Original" code to the "Revised" code (*@\textbf{are clearly focused on fixing vulnerabilities}@*).

Commit Message:
{commit}

Original code snippet (code before changes):
{original}

Revised code snippet (code after changes):
{revised}

Here are the other functions in the same commit:
{context}
\end{lstlisting}
\caption{The Prompt Produces a Score from 0 to 4, Representing the Confidence in Predicting Vulnerability-Fixing Changes.}
\label{fig:range_output}
\end{figure}

\paragraph{Heuristics}
To enhance the capability of LLMs in identifying vulnerability-fixing changes, we applied a heuristic approach. As identified in \cref{sec:motivation}, \emph{Test-Related Changes} significantly contribute to non-vulnerability fixes. Based on this insight, we devised heuristic rules to exclude test-related changes prior to processing the data with LLMs. 

These rules were designed to identify and eliminate test-related code modifications by scrutinizing the naming conventions of the files and functions involved in the changes. We consulted widely-accepted testing frameworks and their naming conventions, such as Pytest for Python, to develop these rules. The comprehensive rules can be found in \cref{fig:heursitcs}.

We offer an example below to demonstrate the procedure. 
\textit{Pytest Naming Conventions:} The file name should commence with the word "test" followed by the name of the file. An underscore symbol "\_" is utilized to separate the terms "test" and the file name for visualization purposes. 
In accordance with this naming convention, we classify code modifications as test-related code changes if the function names contain the word \emph{test}. By implementing these heuristic rules, we effectively filter out test-related changes, enabling the LLMs to concentrate on vulnerability-fixing changes and potentially enhancing their performance in identifying such changes.

\begin{table}[ht]
\centering
\caption{Number of Vulnerable Functions in Curated Datasets}
\label{tb:cleanvul}
\resizebox{0.6\textwidth}{!}{
\begin{tabular}{@{}cccc@{}}
\toprule
Dataset Name & Threshold & Heuristic & Vulnerable Function (\#) \\
\midrule
\multirow{8}{*}{\dataset} & \multirow{2}{*}{1} & \mycmark & 26,518 \\
 & & \myxmark & 29,810 \\
\cmidrule{2-4}
 & \multirow{2}{*}{2} & \mycmark & 16,277 \\
 &  & \myxmark & 18,455 \\
\cmidrule{2-4}
 & \multirow{2}{*}{3} & \mycmark & 8,198 \\
 &  & \myxmark & 9,026 \\
\cmidrule{2-4}
 & \multirow{2}{*}{4} & \mycmark & 6,368 \\
 &  & \myxmark & 7,020 \\
\bottomrule
\end{tabular}
}
\begin{flushleft}
\end{flushleft}
\end{table}

\section{\dataset: Dataset Curation}
\label{sec:curation}

To create a high-quality vulnerability dataset using \tool, we conducted a comprehensive crawl of 127,063 repositories, resulting in the acquisition of 5,352,105 commits. Using the keyword-based approach proposed by Bui et al. \cite{bui2024javavfc}, we identified 43,029 function changes from Vulnerability-Fix Commits (VFCs) from these repositories.
We analyzed these VFCs using \tool, which processed code changes across multiple programming languages: 26,423 Java, 6,591 Python, 5,578 C, 4,000 JavaScript, 312 C\#, and 125 C++ changes.

\tool assigns scores from 0 to 4 to each code change, where 0 indicates no relation to vulnerability fixes and 4 signifies a strong focus on fixing vulnerabilities. Table \ref{tb:cleanvul} presents the distribution of vulnerable functions across these threshold levels in our dataset \dataset, evaluated both with (\mycmark) and without (\myxmark) heuristics. At threshold level 1, we identified 26,518 vulnerable functions with heuristics and 29,810 without. The number of detected functions decreases as the threshold increases: level 2 found 18,455 (\mycmark) and 16,277 (\myxmark) functions; level 3 identified 8,198 (\mycmark) and 9,026 (\myxmark) functions; and level 4 contained 6,368 (\mycmark) and 7,020 (\myxmark) functions.

By varying these threshold values, we created datasets of different quality levels. The dataset with threshold 4 represents the highest confidence in vulnerability-fixing changes, while the dataset with threshold 1 provides broader coverage by including potential vulnerability fixes.

\section{Evaluation Settings}
\label{sec:eva_settings}

This evaluation examines the following research questions (RQs):

\paragraph{RQ1:} \textit{(Efficacy) What is the correctness improvement of \dataset compared to the uncleaned dataset?}\\
\textbf{Rationale:}
After applying \tool to clean the noise in the VFC dataset, it is important to evaluate the cleaned dataset (\dataset) to understand if the percentage of vulnerable functions (\emph{Correctness}) in the dataset has increased and the extent of the improvement.

\paragraph{RQ2:} \textit{(Effectiveness) How does the performance of LLMs fine-tuned on \dataset compare to their performance on other established vulnerability datasets?}\\
\textbf{Rationale:}
According to recent work \cite{ding2024vulnerability}, most vulnerability datasets include 40\% to 75\% of noisy data. In this study, we select the most refined high-quality datasets, including PrimeVul and SVEN, to train machine learning models and perform cross-dataset validation to evaluate the predictive performance of models trained on different datasets. This sheds light on the overall performance of models trained on different datasets, indicating the quality of these datasets.

\paragraph{RQ3:} \textit{(Effectiveness) How does the performance of LLMs fine-tuned on \dataset compare to their performance on the uncleaned dataset?}\\
\textbf{Rationale:}
Apart from cross-comparison with other datasets, we also compare the performance of models trained on our \dataset dataset and uncleaned data to assess the improvement in performance on the original dataset. This helps us understand how much performance is improved after adopting \tool to remove non-vulnerability-fixing changes in VFCs.

\subsection{Comparison with Established Datasets}

Previous studies \cite{ding2024vulnerability,chen2023diversevul} identified only two vulnerability datasets with \emph{Correctness} above 70\%: SVEN \cite{he2023large} and PrimeVul \cite{ding2024vulnerability}. These datasets are comparable to our \dataset, which achieves over 80\% \emph{Correctness} at thresholds 3 and 4.
SVEN, developed through manual analysis, contains 803 functions and achieves 94.0\% \emph{Correctness}. PrimeVul identifies vulnerabilities by matching function names with NVD descriptions and contains 6,968 functions with 86.0\% \emph{Correctness}. However, PrimeVul is limited to NVD-linked vulnerabilities and cannot clean VFCs without NVD entries. We compare the generalizability of models fine-tuned on these established datasets versus our \dataset.

\subsection{Model Choices} 

Our experiments encompass various state-of-the-art LLMs.
We have selected three widely-used encoder-only models, namely RoBERTa \cite{liu2019roberta}, CodeBERT \cite{feng2020codebert}, and GraphCodeBERT \cite{guo2020graphcodebert}, along with three decoder-only models, which consist of two smaller models, GPT-2 \cite{radford2019language} and CodeGPT \cite{lu2021codexglue}, and one larger decoder-only model, CodeLlama \cite{roziere2023code}. 
These models have demonstrated exceptional performance in code-related tasks, making them suitable candidates for our study \cite{xu2022systematic}.

\subsection{Evaluation Metrics}
\label{sec:metrics}

Four popular metrics are adopted in our experiments to assess the performance of the LLMs on tasks such as identifying vulnerability-fixing changes, comparing models trained on different datasets, and more:

\begin{itemize}
    \item \textbf{Accuracy:} This metric measures the overall correctness of the model across all classes. It is defined as the ratio of correctly predicted observations (both true positives and true negatives) to the total number of observations. Our dataset consists of paired samples (vulnerable/benign function pairs), making the classes inherently balanced. Therefore, accuracy is particularly useful for evaluating the performance of models trained on our dataset.
    
    \item \textbf{Precision:} Precision is the ratio of correctly predicted positive observations to the total predicted positives. This metric is crucial when the cost of a false positive is high. In the context of this work, high precision means that most of the changes identified by the model as vulnerability-fixing are indeed correct.

    \item \textbf{Recall:} Recall is the ratio of correctly predicted positive observations to all actual positives. It measures the model's ability to find all relevant cases within a dataset. High recall is important in scenarios where missing an actual positive (failing to identify a true vulnerability-fixing change) can have serious implications.

    \item \textbf{F1-Score:} The F1-score is the harmonic mean of precision and recall. It is a way of combining both precision and recall into a single measure that captures both properties. This metric is particularly useful when you need to balance precision and recall, which often have an inverse relationship.
\end{itemize}

Each of these metrics offers distinct insights into the performance characteristics of the models used in our experiments, allowing us to tailor our model selection and tuning to the specific needs of the application at hand. In addition to these metrics, we also employ the \emph{Correctness} metric to evaluate the quality of the vulnerability dataset. 

\begin{itemize}
    \item \textbf{Correctness}: Correctness is defined as the percentage of genuine vulnerable functions in the vulnerability dataset. This metric is particularly relevant in our study, as it helps assess the effectiveness of our approach in identifying and filtering vulnerability-fixing changes.
\end{itemize}

\subsection{Implementation Details}

For inference with commercial LLMs, we employ the LangChain framework. In order to fine-tune models such as RoBERTa, CodeBERT, GraphCodeBERT, GPT-2, CodeGPT, and CodeLlama, we utilize the Huggingface Transformers library. We set the maximum number of input tokens to 512 to ensure efficient processing.
In terms of training, we configure the number of epochs to 50 and store the optimal checkpoints for later use. To facilitate a seamless transition from pretraining to fine-tuning, we maintain a consistent learning rate that aligns with the LLM's pretraining learning rate.
Our experiments were executed on NVIDIA H100 GPUs, utilizing a server equipped with an Intel(R) Xeon(R) Platinum 8480C CPU and running Ubuntu 22.04.2 LTS as the operating system.

\section{Evaluation Experiments}
\label{sec:eva}

In this section, we report our experimental results and answer the three research questions.

\subsection{RQ1: What is the correctness improvement of \dataset compared to the uncleaned dataset?}
\label{sec:rq3.1}

To evaluate \emph{Correctness} of \dataset, we randomly select a sample of 487 function-level code changes from the collected GitHub VFC dataset and manually analyze them to create a testing dataset. This sample size provides a confidence level of 95\% with a margin of error of ±4.4\% for the full dataset, allowing us to make statistically significant observations. We carefully examine each code change and mark them as either vulnerability-fixing changes or other code changes, establishing a ground truth for our evaluation. 
This evaluation method aligns with previous work \cite{ding2024vulnerability}.
As mentioned in \cref{sec:approach}, \tool's output score ranges from 0 to 4, with 0 indicating no relation to vulnerability fixes and 4 signifying a strong focus on fixing vulnerabilities. We construct four datasets by setting different threshold values. A very clean dataset includes only items scored as 4, representing the highest confidence in vulnerability-fixing changes. In contrast, a more extensive dataset retains items scored 1 or higher, encompassing potential vulnerability-fixing changes. The results are presented in \cref{tb:vfc}.

\begin{table}[ht]
\centering
\caption{Comparison of Accuracy Across Existing Vulnerability Datasets With Our Cleaned Dataset}
\label{tb:vfc}
\resizebox{0.9\textwidth}{!}{
\begin{tabular}{@{}ccccccc@{}}
\toprule
Dataset & Threshold & Heuristic & Vulnerable Func. & Ori. Corr. & Corr. & Improvement \\ 
\midrule
\multirow{8}{*}{\dataset} & \multirow{2}{*}{1} & \mycmark & 26,518 & 28.7\% &  43.1\% & +50.1\% \\
 & & \myxmark & 29,810 & 28.7\% & 37.5\% & +30.6\% \\
\cmidrule{2-7}
 & \multirow{2}{*}{2} & \mycmark & 16,277 & 28.7\% & 49.4\% & +72.1\% \\
 &  & \myxmark & 18,455 & 28.7\% & 57.7\% & +101.0\% \\
\cmidrule{2-7}
 & \multirow{2}{*}{3} & \mycmark & 8,198 & 28.7\% & 90.6\% & +215.6\% \\
 &  & \myxmark & 9,026 & 28.7\% & 76.5\% & +166.5\% \\
\cmidrule{2-7}
 & \multirow{2}{*}{4} & \mycmark & 6,368 & 28.7\% & 97.3\% & +239.0\% \\
 &  & \myxmark & 7,020 & 28.7\% & 78.0\% & +171.7\% \\
\midrule
SVEN~\cite{he2023large} & - & - & 803 & - & 94.0\%$^*$ & - \\
PrimeVul~\cite{ding2024vulnerability} & - & - & 6,968 & - & 86.0\%$^*$ & - \\
DiverseVul~\cite{chen2023diversevul} & - & - & 18,945 & - & 60.0\%$^\dagger$ & - \\
CVEFixes~\cite{bhandari2021cvefixes} & - & - & 5,495 & - & 51.7\%$^\dagger$ & - \\
CrossVul~\cite{nikitopoulos2021crossvul} & - & - & 5,877 & - & 47.8\%$^\dagger$ & - \\
VulnPatchPairs~\cite{risse2023limits} & - & - & 13,100 & - & 36.0\%$^*$ & - \\
BigVul~\cite{fan2020ac} & - & - & 11,823 & - & 25.0\%$^\dagger$ & - \\
CodeXGLUE~\cite{zhou2019devign,lu2021codexglue} & - & - & 23,355 & - & 24.0\%$^*$ & - \\
\bottomrule
\end{tabular}
}
\begin{flushleft}
\small
$^*$ Refers to results in Ding et al.~\cite{ding2024vulnerability}.\\
$\dagger$ Refers to results in Chen et al.~\cite{chen2023diversevul}.
\end{flushleft}
\end{table}

Upon applying \tool, the \emph{Correctness} (defined in \cref{sec:metrics} as the percentage of genuine vulnerable functions in the vulnerability dataset) improves from 28.7\% to a range of 37.5\% to 97.3\% on the test sample, with improvements spanning from 30.6\% to 239.0\%. Notably, the heuristic approach enhances \emph{Correctness} across different thresholds. When the threshold is set to 1, the \emph{Correctness} increases from 28.7\% to 43.1\% with the heuristic, resulting in a 50.1\% improvement. Without the heuristic, the \emph{Correctness} reaches 37.5\%, yielding an 30.6\% improvement. For thresholds 2, 3, and 4, the heuristic approach yields \emph{Correctness} improvements of 101.0\%, 215.6\%, and 239.0\%, respectively. Thresholds 3 and 4 demonstrate particularly high \emph{Correctness} levels.

\paragraph{Comparison with Other Datasets}
Most existing datasets exhibit low \emph{Correctness} levels since they treat all changes in VFCs as vulnerability-fixing changes. Only SVEN~\cite{he2023large} and PrimeVul~\cite{ding2024vulnerability} verify this property and achieve good \emph{Correctness}. However, SVEN is limited in size, containing only 803 vulnerable functions due to its reliance on manual analysis. PrimeVul, while larger, requires NVD links, making it unsuitable for VFCs without associated NVD entries.
The enhanced \emph{Correctness} using \tool is competitive with these established datasets. At a threshold of 3, \tool achieves a \emph{Correctness} rate of 90.6\%, comparable to SVEN~\cite{he2023large} (94.0\%) and PrimeVul~\cite{ding2024vulnerability} (86.0\%). This achievement is particularly noteworthy given that our \dataset dataset is derived from GitHub and operates at a larger scale, in contrast to others that primarily rely on NVD data. As such, our \dataset dataset serves as a crucial complement, broadening the scope and application of vulnerability datasets in real-world scenarios.

\begin{framed}
\noindent With a threshold of 3, we obtain a dataset of \textbf{8,198} vulnerability-fixing changes, achieving a \emph{Correctness} of \textbf{90.6\%} on the testing sample. Increasing the threshold to 4 results in \textbf{6,368} vulnerability-fixing changes with a perfect \emph{Correctness} of \textbf{97.3\%}. These results match the \emph{Correctness} levels of existing high-quality datasets SVEN~\cite{he2023large} (94.0\%) and PrimeVul~\cite{ding2024vulnerability} (86.0\%), while providing more samples compared to SVEN (803) and PrimeVul (6,968).
\end{framed}

\subsection{RQ2: How does the performance of LLMs fine-tuned on \dataset compare to their performance on other established vulnerability datasets?}
\label{sec:rq3.2}

\begin{table}
\centering
\caption{Comparison with Different LLMs Fine-Tuned on \dataset (\textbf{Incorporating All Programming Languages}) and Other Datasets}
\label{tb:data_comparison_all}
\resizebox{0.75\textwidth}{!}{
\begin{tabular}{>{\centering\arraybackslash}p{2.8cm}|>{\centering\arraybackslash}p{2.2cm}|>{\centering\arraybackslash}p{2.2cm}|>{\centering\arraybackslash}p{1.2cm}>{\centering\arraybackslash}p{1.2cm}>{\centering\arraybackslash}p{1.2cm}>{\centering\arraybackslash}p{1.2cm}}
\toprule
Model & Train & Test & Acc (\%) & Pre (\%) & Rec (\%) & F1 (\%) \\
\midrule
\multirow{9}{*}{\makecell{RoBERTa\\125M\\Encoder-Only}} & \cellcolor{color-b!25}\dataset & \cellcolor{color-b!25}\dataset & \cellcolor{color-b!25}65.83 & \cellcolor{color-b!25}64.55 & \cellcolor{color-b!25}71.17 & \cellcolor{color-b!25}67.56 \\
 & & PrimeVul & 56.26 & 55.79 & 60.80 & 58.18 \\
 & & SVEN & 60.97 & 57.91 & 80.20 & 67.23 \\
\cmidrule{2-7}
 & \cellcolor{color-b!25}PrimeVul & \dataset & 52.53 & 52.01 & 65.88 & 57.85 \\
 & & \cellcolor{color-b!25}PrimeVul & \cellcolor{color-b!25}51.63 & \cellcolor{color-b!25}51.77 & \cellcolor{color-b!25}52.17 & \cellcolor{color-b!25}51.15 \\
 & & SVEN & 49.72 & 49.75 & 75.17 & 59.34 \\
\cmidrule{2-7}
 & \cellcolor{color-b!25}SVEN & \dataset & 50.98 & 51.28 & 51.33 & 50.10 \\
 & & PrimeVul & 52.32 & 52.72 & 52.83 & 48.98 \\
 & & \cellcolor{color-b!25}SVEN & \cellcolor{color-b!25}75.00 & \cellcolor{color-b!25}77.90 & \cellcolor{color-b!25}70.91 & \cellcolor{color-b!25}73.77 \\
\midrule
\multirow{9}{*}{\makecell{CodeBERT\\125M\\Encoder-Only}} & \cellcolor{color-b!25}\dataset & \cellcolor{color-b!25}\dataset & \cellcolor{color-b!25}68.10 & \cellcolor{color-b!25}66.98 & \cellcolor{color-b!25}71.52 & \cellcolor{color-b!25}69.06 \\
 & & PrimeVul & 58.09 & 58.06 & 61.85 & 59.39 \\
 & & SVEN & 64.87 & 60.99 & 85.14 & 70.87 \\
\cmidrule{2-7}
 & \cellcolor{color-b!25}PrimeVul & \dataset & 54.97 & 53.92 & 68.29 & 60.25 \\
 & & \cellcolor{color-b!25}PrimeVul & \cellcolor{color-b!25}56.61 & \cellcolor{color-b!25}56.29 & \cellcolor{color-b!25}58.44 & \cellcolor{color-b!25}57.03 \\
 & & SVEN & 54.94 & 53.42 & 77.48 & 63.23 \\
\cmidrule{2-7}
 & \cellcolor{color-b!25}SVEN & \dataset & 52.93 & 51.98 & 79.24 & 62.67 \\
 & & PrimeVul & 53.57 & 52.81 & 74.80 & 61.33 \\
 & & \cellcolor{color-b!25}SVEN & \cellcolor{color-b!25}81.32 & \cellcolor{color-b!25}77.25 & \cellcolor{color-b!25}88.99 & \cellcolor{color-b!25}82.64 \\
\midrule
\multirow{9}{*}{\makecell{GraphCodeBERT\\125M\\Encoder-Only}} & \cellcolor{color-b!25}\dataset & \cellcolor{color-b!25}\dataset & \cellcolor{color-b!25}68.96 & \cellcolor{color-b!25}66.09 & \cellcolor{color-b!25}78.11 & \cellcolor{color-b!25}71.57 \\
 & & PrimeVul & 54.65 & 53.75 & 66.87 & 59.59 \\
 & & SVEN & 62.07 & 58.03 & 87.91 & 69.88 \\
\cmidrule{2-7}
 & \cellcolor{color-b!25}PrimeVul & \dataset & 54.19 & 52.89 & 77.08 & 62.72 \\
 & & \cellcolor{color-b!25}PrimeVul & \cellcolor{color-b!25}57.19 & \cellcolor{color-b!25}55.50 & \cellcolor{color-b!25}72.39 & \cellcolor{color-b!25}62.81 \\
 & & SVEN & 55.75 & 54.43 & 70.86 & 61.57 \\
\cmidrule{2-7}
 & \cellcolor{color-b!25}SVEN & \dataset & 54.74 & 53.22 & 79.36 & 63.69 \\
 & & PrimeVul & 52.50 & 52.01 & 64.94 & 57.76 \\
 & & \cellcolor{color-b!25}SVEN & \cellcolor{color-b!25}81.86 & \cellcolor{color-b!25}79.60 & \cellcolor{color-b!25}85.71 & \cellcolor{color-b!25}82.54 \\
\midrule
\multirow{9}{*}{\makecell{GPT-2\\124M\\Decoder-Only}} & \cellcolor{color-b!25}\dataset & \cellcolor{color-b!25}\dataset & \cellcolor{color-b!25}61.03 & \cellcolor{color-b!25}61.63 & \cellcolor{color-b!25}58.47 & \cellcolor{color-b!25}59.94 \\
 & & PrimeVul & 53.13 & 53.01 & 60.43 & 56.03 \\
 & & SVEN & 54.37 & 53.61 & 66.12 & 59.17 \\
\cmidrule{2-7}
 & \cellcolor{color-b!25}PrimeVul & \dataset & 51.75 & 51.75 & 51.12 & 51.32 \\
 & & \cellcolor{color-b!25}PrimeVul & \cellcolor{color-b!25}51.26 & \cellcolor{color-b!25}51.42 & \cellcolor{color-b!25}45.28 & \cellcolor{color-b!25}48.14 \\
 & & SVEN & 55.74 & 56.54 & 49.18 & 52.58 \\
\cmidrule{2-7}
 & \cellcolor{color-b!25}SVEN & \dataset & 52.36 & 51.92 & 67.41 & 58.38 \\
 & & PrimeVul & 51.50 & 51.44 & 60.97 & 54.67 \\
 & & \cellcolor{color-b!25}SVEN & \cellcolor{color-b!25}78.69 & \cellcolor{color-b!25}78.34 & \cellcolor{color-b!25}80.33 & \cellcolor{color-b!25}78.95 \\
\midrule
\multirow{9}{*}{\makecell{CodeGPT\\124M\\Decoder-Only}} & \cellcolor{color-b!25}\dataset & \cellcolor{color-b!25}\dataset & \cellcolor{color-b!25}66.81 & \cellcolor{color-b!25}67.32 & \cellcolor{color-b!25}65.57 & \cellcolor{color-b!25}66.38 \\
 & & PrimeVul & 56.25 & 57.40 & 49.69 & 53.11 \\
 & & SVEN & 56.83 & 54.97 & 75.96 & 63.76 \\
 \cmidrule{2-7}
 & \cellcolor{color-b!25}PrimeVul & \dataset & 52.62 & 52.08 & 63.79 & 57.14 \\
 & & \cellcolor{color-b!25}PrimeVul & \cellcolor{color-b!25}53.02 & \cellcolor{color-b!25}54.20 & \cellcolor{color-b!25}39.16 & \cellcolor{color-b!25}45.44 \\
 & & SVEN & 51.91 & 51.57 & 60.11 & 53.87 \\
\cmidrule{2-7}
 & \cellcolor{color-b!25}SVEN & \dataset & 52.91 & 52.71 & 56.49 & 54.42 \\
 & & PrimeVul & 55.20 & 56.75 & 43.12 & 48.11 \\
 & & \cellcolor{color-b!25}SVEN & \cellcolor{color-b!25}79.78 & \cellcolor{color-b!25}83.38 & \cellcolor{color-b!25}74.86 & \cellcolor{color-b!25}78.63 \\
\midrule
\multirow{9}{*}{\makecell{CodeLlama\\7B\\Decoder-Only}} & \cellcolor{color-b!25}\dataset & \cellcolor{color-b!25}\dataset & \cellcolor{color-b!25}60.98 & \cellcolor{color-b!25}60.51 & \cellcolor{color-b!25}64.67 & \cellcolor{color-b!25}62.46 \\
 & & PrimeVul & 53.46 & 53.74 & 45.92 & 49.05 \\
 & & SVEN & 53.14 & 52.63 & 64.27 & 57.83 \\
\cmidrule{2-7}
 & \cellcolor{color-b!25}PrimeVul & \dataset & 51.59 & 48.73 & 44.17 & 40.06 \\
 & & \cellcolor{color-b!25}PrimeVul & \cellcolor{color-b!25}52.82 & \cellcolor{color-b!25}51.90 & \cellcolor{color-b!25}38.88 & \cellcolor{color-b!25}39.04 \\
 & & SVEN & 52.56 & 68.36 & 48.16 & 41.37 \\
\cmidrule{2-7}
 & \cellcolor{color-b!25}SVEN & \dataset & 50.53 & 51.21 & 49.36 & 45.60 \\
 & & PrimeVul & 50.27 & 49.43 & 49.86 & 42.19 \\
 & & \cellcolor{color-b!25}SVEN & \cellcolor{color-b!25}51.66 & \cellcolor{color-b!25}52.19 & \cellcolor{color-b!25}53.65 & \cellcolor{color-b!25}48.40 \\
\bottomrule
\end{tabular}
}
\end{table}

To evaluate the effectiveness of our \dataset dataset, we conducted two experiments: 1) we focused on fine-tuning various LLMs on all programming languages in the \dataset dataset and other high-quality datasets; 2) since our \dataset primarily contains vulnerable code in Java, we trained various LLMs on Java only in the \dataset and compared them with models fine-tuned on other datasets to assess cross-language and cross-dataset performance. 

Our \dataset features a vulnerability score ranging from 0 to 4. To test the highest quality dataset we could obtain, we selected a threshold of 4. As mentioned in \cref{sec:rq3.1}, this process resulted in a collection of 6,368 vulnerable and benign function pairs. To facilitate comparison with two other high-quality datasets, SVEN~\cite{he2023large} and PrimeVul~\cite{ding2024vulnerability}, which exhibit high \emph{Correctness} levels exceeding 85\%, we also employed a balanced dataset setting for training and testing to eliminate other factors. It is important to note that PrimeVul contains 6,968 pairs of code changes exclusively in C and C++ code, while SVEN consists of 803 pairs of code changes, with equal parts C and C++ code and Python code. We partitioned the dataset into training, testing, and validation sets using a 7:1.5:1.5 ratio and subsequently compared their cross-dataset performance. The evaluation metrics employed were accuracy, precision, recall, and F1-score, which are presented for different LLMs trained on one dataset and tested on others, as illustrated in Table \ref{tb:data_comparison}.

\paragraph{Intra-Dataset Performance Fine-Tuned on All Languages}
Considering that the datasets used for training and testing are balanced in this study, we primarily compare and report accuracy. The results indicate that models trained on \dataset generally exhibit good performance on the same dataset. For example, GraphCodeBERT achieved the best intra-dataset performance for all three datasets. Specifically, GraphCodeBERT fine-tuned and tested on \dataset achieved an accuracy of 68.96\%, 57.19\% on PrimeVul, and 81.86\% on SVEN. The accuracy is higher than PrimeVul (68.96\% vs 57.19\%), but lower than SVEN (81.86\%). This discrepancy might be due to the low diversity of the SVEN dataset, which contains only around 800 vulnerability functions across 10 CWE vulnerabilities. Another reason could be the diversity in programming languages, as \dataset includes six different languages: Java, Python, C, C++, C\#, and JavaScript, while SVEN contains only Python, C, and C++. Interestingly, larger models such as CodeLlama-7B achieved the lowest accuracy of 60.98\% compared to the smaller models, which could be attributed to a form of underfitting, possibly due to insufficient training data or inadequate model architecture for capturing the nuances of the diverse programming languages and vulnerabilities.

\paragraph{Inter-Dataset Generalization Fine-Tuned on All Languages}
Our dataset \dataset demonstrated excellent performance during cross-dataset generalization experiments. Specifically, regarding generalization to PrimeVul, when fine-tuned on our dataset and tested on PrimeVul, the best accuracy is 58.09\% with CodeBERT, which is even higher than solely fine-tuning and testing on PrimeVul with the best accuracy of 57.19\%. The best model that fine-tuned on SVEN and tested on PrimeVul is CodeGPT with an accuracy of 55.20\%, which is lower than fine-tuning and testing on PrimeVul with the best accuracy of 57.19\%. This shows that models fine-tuned on our dataset achieved the best performance on PrimeVul compared to models fine-tuned on PrimeVul or SVEN.

Regarding generalization to SVEN, when fine-tuned on \dataset and tested on SVEN, the best accuracy is 64.87\% with CodeBERT, which is higher than the models fine-tuned on PrimeVul and tested on SVEN with an accuracy of 55.75\% with GraphCodeBERT. Although both are lower than only training and testing on SVEN, with an accuracy of 81.86\% with GraphCodeBERT, our dataset demonstrated impressive performance in generalizing knowledge to an unknown dataset.

In terms of training on other two datasets and testing on \dataset, we can notice that the best accuracy for SVEN is 54.74\% with GraphCodeBERT, and the best accuracy for PrimeVul is 54.97\% with CodeBERT. Both accuracies are lower than the accuracy trained and tested on our dataset (68.96\%). This shows that our dataset might be more diverse than PrimeVul and SVEN.

\begin{table}
\centering
\caption{Comparison with Different LLMs Fine-Tuned on \dataset (\textbf{Java Only}) and Other Datasets}
\label{tb:data_comparison}
\resizebox{0.75\textwidth}{!}{
\begin{tabular}{>{\centering\arraybackslash}p{2.8cm}|>{\centering\arraybackslash}p{2.2cm}|>{\centering\arraybackslash}p{2.2cm}|>{\centering\arraybackslash}p{1.2cm}>{\centering\arraybackslash}p{1.2cm}>{\centering\arraybackslash}p{1.2cm}>{\centering\arraybackslash}p{1.2cm}}
\toprule
Model & Train & Test & Acc (\%) & Pre (\%) & Rec (\%) & F1 (\%) \\
\midrule
\multirow{9}{*}{\makecell{RoBERTa\\125M\\Encoder-Only}} & \cellcolor{color-b!25}\dataset & \cellcolor{color-b!25}\dataset & \cellcolor{color-b!25}66.96 & \cellcolor{color-b!25}65.60 & \cellcolor{color-b!25}71.30 & \cellcolor{color-b!25}68.33 \\
 & & PrimeVul & 50.66 & 50.47 & 70.61 & 58.87 \\
 & & SVEN & 51.64 & 51.28 & 65.57 & 57.55 \\
\cmidrule{2-7}
 & \cellcolor{color-b!25}PrimeVul & \dataset & 52.61 & 51.79 & 75.65 & 61.48 \\
 & & \cellcolor{color-b!25}PrimeVul & \cellcolor{color-b!25}53.51 & \cellcolor{color-b!25}53.23 & \cellcolor{color-b!25}57.89 & \cellcolor{color-b!25}55.46 \\
 & & SVEN & 51.64 & 51.00 & 83.61 & 63.35 \\
\cmidrule{2-7}
 & \cellcolor{color-b!25}SVEN & \dataset & 53.48 & 53.17 & 58.26 & 55.60 \\
 & & PrimeVul & 53.73 & 53.36 & 59.21 & 56.13 \\
 & & \cellcolor{color-b!25}SVEN & \cellcolor{color-b!25}76.23 & \cellcolor{color-b!25}75.00 & \cellcolor{color-b!25}78.69 & \cellcolor{color-b!25}76.80 \\
\midrule
\multirow{9}{*}{\makecell{CodeBERT\\125M\\Encoder-Only}} & \cellcolor{color-b!25}\dataset & \cellcolor{color-b!25}\dataset & \cellcolor{color-b!25}73.04 & \cellcolor{color-b!25}68.53 & \cellcolor{color-b!25}85.22 & \cellcolor{color-b!25}75.97 \\
 & & PrimeVul & 52.41 & 51.80 & 69.30 & 59.29 \\
 & & SVEN & 50.82 & 50.44 & 93.44 & 65.52 \\
\cmidrule{2-7}
 & \cellcolor{color-b!25}PrimeVul & \dataset & 53.91 & 52.98 & 69.57 & 60.15 \\
 & & \cellcolor{color-b!25}PrimeVul & \cellcolor{color-b!25}54.17 & \cellcolor{color-b!25}54.15 & \cellcolor{color-b!25}54.39 & \cellcolor{color-b!25}54.27 \\
 & & SVEN & 57.38 & 54.95 & 81.97 & 65.79 \\
\cmidrule{2-7}
 & \cellcolor{color-b!25}SVEN & \dataset & 54.35 & 54.24 & 55.65 & 54.94 \\
 & & PrimeVul & 53.51 & 52.58 & 71.49 & 60.59 \\
 & & \cellcolor{color-b!25}SVEN & \cellcolor{color-b!25}85.25 & \cellcolor{color-b!25}85.25 & \cellcolor{color-b!25}85.25 & \cellcolor{color-b!25}85.25 \\
\midrule
\multirow{9}{*}{\makecell{GraphCodeBERT\\125M\\Encoder-Only}} & \cellcolor{color-b!25}\dataset & \cellcolor{color-b!25}\dataset & \cellcolor{color-b!25}74.78 & \cellcolor{color-b!25}71.11 & \cellcolor{color-b!25}83.48 & \cellcolor{color-b!25}76.80 \\
 & & PrimeVul & 52.19 & 51.48 & 76.32 & 61.48 \\
 & & SVEN & 54.10 & 52.48 & 86.89 & 65.43 \\
\cmidrule{2-7}
 & \cellcolor{color-b!25}PrimeVul & \dataset & 50.43 & 50.30 & 73.04 & 59.57 \\
 & & \cellcolor{color-b!25}PrimeVul & \cellcolor{color-b!25}55.26 & \cellcolor{color-b!25}54.92 & \cellcolor{color-b!25}58.77 & \cellcolor{color-b!25}56.78 \\
 & & SVEN & 59.84 & 56.98 & 80.33 & 66.67 \\
\cmidrule{2-7}
 & \cellcolor{color-b!25}SVEN & \dataset & 51.30 & 51.22 & 54.78 & 52.94 \\
 & & PrimeVul & 52.63 & 52.13 & 64.47 & 57.65 \\
 & & \cellcolor{color-b!25}SVEN & \cellcolor{color-b!25}83.61 & \cellcolor{color-b!25}85.96 & \cellcolor{color-b!25}80.33 & \cellcolor{color-b!25}83.05 \\
\midrule
\multirow{9}{*}{\makecell{GPT-2\\124M\\Decoder-Only}} & \cellcolor{color-b!25}\dataset & \cellcolor{color-b!25}\dataset & \cellcolor{color-b!25}71.30 & \cellcolor{color-b!25}71.30 & \cellcolor{color-b!25}71.30 & \cellcolor{color-b!25}71.30 \\
 & & PrimeVul & 53.29 & 52.49 & 69.30 & 59.74 \\
 & & SVEN & 59.02 & 56.79 & 75.41 & 64.79 \\
\cmidrule{2-7}
 & \cellcolor{color-b!25}PrimeVul & \dataset & 50.00 & 50.00 & 50.43 & 50.22 \\
 & & \cellcolor{color-b!25}PrimeVul & \cellcolor{color-b!25}52.63 & \cellcolor{color-b!25}53.23 & \cellcolor{color-b!25}43.42 & \cellcolor{color-b!25}47.83 \\
 & & SVEN & 56.56 & 56.06 & 60.66 & 58.27 \\
\cmidrule{2-7}
 & \cellcolor{color-b!25}SVEN & \dataset & 54.78 & 52.97 & 85.22 & 65.33 \\
 & & PrimeVul & 53.73 & 54.03 & 50.00 & 51.94 \\
 & & \cellcolor{color-b!25}SVEN & \cellcolor{color-b!25}80.33 & \cellcolor{color-b!25}79.37 & \cellcolor{color-b!25}81.97 & \cellcolor{color-b!25}80.65 \\
\midrule
\multirow{9}{*}{\makecell{CodeGPT\\124M\\Decoder-Only}} & \cellcolor{color-b!25}\dataset & \cellcolor{color-b!25}\dataset & \cellcolor{color-b!25}73.36 & \cellcolor{color-b!25}73.17 & \cellcolor{color-b!25}73.77 & \cellcolor{color-b!25}73.47 \\
 & & PrimeVul & 55.13 & 54.51 & 61.97 & 58.00 \\
 & & SVEN & 52.46 & 51.58 & 80.33 & 62.82 \\
\cmidrule{2-7}
 & \cellcolor{color-b!25}PrimeVul & \dataset & 53.28 & 52.11 & 81.15 & 63.46 \\
 & & \cellcolor{color-b!25}PrimeVul & \cellcolor{color-b!25}54.49 & \cellcolor{color-b!25}57.34 & \cellcolor{color-b!25}35.04 & \cellcolor{color-b!25}43.50 \\
 & & SVEN & 50.00 & 50.00 & 86.89 & 63.47 \\
\cmidrule{2-7}
 & \cellcolor{color-b!25}SVEN & \dataset & 50.00 & 50.00 & 76.23 & 60.39 \\
 & & PrimeVul & 52.99 & 52.82 & 55.98 & 54.36 \\
 & & \cellcolor{color-b!25}SVEN & \cellcolor{color-b!25}82.79 & \cellcolor{color-b!25}83.33 & \cellcolor{color-b!25}81.97 & \cellcolor{color-b!25}82.64 \\
\midrule
\multirow{9}{*}{\makecell{CodeLlama\\7B\\Decoder-Only}} & \cellcolor{color-b!25}\dataset & \cellcolor{color-b!25}\dataset & \cellcolor{color-b!25}70.21 & \cellcolor{color-b!25}70.92 & \cellcolor{color-b!25}68.49 & \cellcolor{color-b!25}69.69 \\
 & & PrimeVul & 52.50 & 53.02 & 43.85 & 48.00 \\
 & & SVEN & 51.52 & 51.52 & 51.52 & 51.52 \\
\cmidrule{2-7}
 & \cellcolor{color-b!25}PrimeVul & \dataset & 50.68 & 62.50 & 3.42 & 6.49 \\
 & & \cellcolor{color-b!25}PrimeVul & \cellcolor{color-b!25}49.81 & \cellcolor{color-b!25}47.37 & \cellcolor{color-b!25}3.46 & \cellcolor{color-b!25}6.45 \\
 & & SVEN & 50.76 & 100.00 & 1.52 & 2.99 \\
\cmidrule{2-7}
 & \cellcolor{color-b!25}SVEN & \dataset & 49.32 & 49.02 & 34.25 & 40.32 \\
 & & PrimeVul & 50.00 & 50.00 & 69.62 & 58.20 \\
 & & \cellcolor{color-b!25}SVEN & \cellcolor{color-b!25}50.76 & \cellcolor{color-b!25}50.65 & \cellcolor{color-b!25}59.09 & \cellcolor{color-b!25}54.55 \\
\bottomrule
\end{tabular}
}
\end{table}

\paragraph{Intra-Dataset Performance Fine-Tuned Exclusively on Java}
Similar to the models fine-tuned on all programming languages on \dataset, the results fine-tuned exclusively on Java on \dataset also demonstrate very good intra-dataset performance. For instance, CodeGPT achieved an accuracy of 73.36\%, outperforming the intra-dataset performance of PrimeVul (54.49\%) but still lower than the accuracy of SVEN (82.79\%). Comparing different LLMs, we observed that GraphCodeBERT is the best-performing encoder-only model with an accuracy of 74.78\%, while CodeGPT is the top-performing decoder-only model with an accuracy of 73.36\%. Similarly, larger models such as CodeLlama-7B achieved a lower accuracy of 70.21\% compared to these smaller models. Moreover, the intra-dataset performance fine-tuned exclusively on Java is improved in comparison with fine-tuning on all programming languages on \dataset, as the best accuracy increased from 68.96\% to 74.78\%.

\paragraph{Inter-Dataset Generalization Fine-Tuned Exclusively on Java}
In cross-dataset testing, when training GPT-2 on \dataset on Java and testing on SVEN, the accuracy reached 59.02\%, which is comparable to the best accuracy achieved by training GraphCodeBERT on PrimeVul and testing on SVEN (59.84\%). However, this is still lower than the 85.25\% accuracy obtained when training and testing CodeBERT on SVEN. This finding indicates that \dataset (Java only) and PrimeVul exhibit similar performance when generalizing to SVEN. However, considering that PrimeVul contains only C++ and C code and SVEN contains almost half of C++ and C vulnerable functions, and that models are fine-tuned on \dataset only on Java code, the generalizability of \dataset (Java only) shows comparable performance across languages rather than not cross languages from PrimeVul to SVEN. This demonstrates impressive cross-language performance of \dataset.

Regarding generalization to PrimeVul, training CodeGPT on \dataset on Java and testing on PrimeVul resulted in an accuracy of 55.13\%, which is very close to the best performance achieved by training and testing GraphCodeBERT on PrimeVul (55.26\%). Conversely, when training on PrimeVul and testing on \dataset on Java, the best performance was 53.91\% for CodeBERT, which is considerably lower than the 74.78\% accuracy achieved when training and testing on \dataset (Java only). Additionally, the highest accuracy obtained when training on SVEN and testing on PrimeVul was 53.73\% using GPT-2 or RoBERTa, which is lower than the results from training on \dataset on Java and testing on PrimeVul. This evidence highlights the high quality of our dataset, suggesting that \dataset (Java only) may contain more comprehensive or diverse examples of vulnerability-fixing changes than PrimeVul and SVEN, and that the knowledge from training on \dataset can be generalized to PrimeVul but not vice versa.

In terms of training on SVEN and testing on \dataset (Java only), the best accuracy achieved was 54.78\% using GPT-2. This is similar to the results obtained when training on SVEN and testing on PrimeVul (53.73\%), indicating that the diversity of \dataset and PrimeVul could be much higher than that of SVEN.

\begin{framed}
\noindent Models fine-tuned on \dataset demonstrate \textbf{much better generalization capabilities comparable to those fine-tuned on PrimeVul} when tested on SVEN (64.87\% vs 55.75\%). Additionally, these models exhibit \textbf{superior generalization abilities when tested on PrimeVul} compared to models trained on SVEN (58.09\% vs 55.20\%). Remarkably, the accuracy achieved by models trained on \dataset and tested on PrimeVul is even better than those trained and tested solely on PrimeVul (\textbf{58.09\% vs 57.19\%}), highlighting the effectiveness and robustness of \dataset in model fine-tuning.
\end{framed}

\subsection{RQ3: How does the performance of LLMs fine-tuned on \dataset compare to their performance on the uncleaned dataset?}

To further assess the effectiveness of \tool in cleaning noisy data, we train the same LLMs on both \dataset and the uncleaned dataset, and test their performance on these datasets as well as two other high-quality datasets, PrimeVul and SVEN. Since GraphCodeBERT demonstrates the best generalization ability when trained on \dataset in \cref{sec:rq3.2}, we train GraphCodeBERT on the uncleaned dataset again and compare the performance.

\begin{table}[ht]
\centering
\caption{Comparison of \dataset and Uncleaned Dataset Performance Across Other Datasets}
\label{tb:comparison_with_uncleaned_2}
\resizebox{0.95\textwidth}{!}{
\begin{tabular}{>{\centering\arraybackslash}p{2.8cm}|>{\centering\arraybackslash}p{2.8cm}|>{\centering\arraybackslash}p{2.8cm}|>{\centering\arraybackslash}p{1.2cm}>{\centering\arraybackslash}p{1.2cm}>{\centering\arraybackslash}p{1.2cm}>{\centering\arraybackslash}p{1.2cm}}
\toprule
Model & Train & Test & Acc (\%) & Pre (\%) & Rec (\%) & F1 (\%) \\
\midrule
\multirow{8}{*}{\makecell{GraphCodeBERT\\125M\\Encoder-Only}} & \cellcolor{color-b!25}\dataset & \cellcolor{color-b!25}\dataset & \cellcolor{color-b!25}68.96 & \cellcolor{color-b!25}66.09 & \cellcolor{color-b!25}78.11 & \cellcolor{color-b!25}71.57 \\
& & PrimeVul & 54.65 & 53.75 & 66.87 & 59.59 \\
& & SVEN & 62.07 & 58.03 & 87.91 & 69.88 \\
& & Uncleaned Data & 56.63 & 54.49 & 80.85 & 65.09 \\
\cmidrule{2-7}
& \cellcolor{color-b!25}Uncleaned Data & \dataset & 59.90 & 60.63 & 48.17 & 50.42 \\
& & PrimeVul & 52.42 & 68.41 & 47.22 & 40.50 \\
& & SVEN & 52.19 & 51.45 & 62.46 & 54.33 \\
& & \cellcolor{color-b!25}Uncleaned Data & \cellcolor{color-b!25}55.23 & \cellcolor{color-b!25}55.98 & \cellcolor{color-b!25}43.98 & \cellcolor{color-b!25}46.32 \\
\bottomrule
\end{tabular}
}
\end{table}

The results are presented in \cref{tb:comparison_with_uncleaned_2}. When training and testing on the same dataset, GraphCodeBERT trained on \dataset achieves a higher accuracy compared to the uncleaned dataset, with accuracies of 68.96\% and 55.23\%, and F1-scores of 71.57\% and 46.32\%, respectively. Notably, when trained on \dataset and tested on unclean data, the accuracy reaches 56.63\% and the F1-score is 65.09\%, which is higher than training and testing both on uncleaned data. Furthermore, when trained on uncleaned data and tested on \dataset, the accuracy is 59.90\% and the F1-score is 50.42\%, which is considerably lower than the accuracy of 68.96\% and F1-score of 71.57\% when trained and tested on \dataset. 

Upon examining the performance of training on \dataset and uncleaned data and testing on other high-quality datasets, the differences become even more pronounced. \dataset achieves 54.65\% accuracy and 59.59\% F1-score on PrimeVul, and 62.07\% accuracy and 69.88\% F1-score on SVEN, while the uncleaned dataset only attains accuracies of 52.42\% and 52.19\%, and F1-scores of 40.50\% and 54.33\% on PrimeVul and SVEN, respectively.

\begin{framed}
\noindent \textbf{Training on the \dataset dataset improves accuracy and F1-score when tested on the same dataset and other high-quality datasets} (PrimeVul and SVEN) \textbf{compared to using uncleaned data}. The accuracy rates are 68.96\% versus 55.23\%, 54.65\% versus 52.42\%, and 62.07\% versus 52.19\%, respectively. Similarly, the F1-scores are 71.57\% versus 46.32\%, 59.59\% versus 40.50\%, and 69.88\% versus 54.33\%, respectively, demonstrating the benefits of training on cleaned data.
\end{framed}

\section{Additional Analyses}
\label{sec:analysis}

Beyond the primary evaluation discussed in \cref{sec:eva}, we conducted additional analyses to gain deeper insights into \tool's performance. Our additional investigations include: 1) an ablation study examining the impact of input combinations, the heuristic module, and the decision to use a 0-4 rating scale rather than binary output, and 2) a sensitivity analysis across different LLMs.

\subsection{Sensitivity Analyses}
\label{sec:sensitive}

\paragraph{Manual Analysis}
To evaluate the performance of different LLMs, as no evaluation dataset existed for this task, and no prior studies had proposed automatic approaches for this, we curated a test dataset manually. We conducted a detailed manual analysis involving multiple rounds:

\begin{itemize}
    \item In the \textbf{first round}, we clarified our objectives and engaged five researchers to analyze a batch of previously studied VFCs \cite{bui2024javavfc}. Out of these five researchers, two were assigned to independently analyze each function change. We examined a total of 125 function changes in 50 VFCs, categorizing each change as a vulnerability-fixing change or not, while taking into account the function changes and commit messages. The independent analyses conducted by the two researchers from the group of five were later used to calculate Kappa's agreement.
    \item In the \textbf{second and third rounds}, the same five researchers analyzed another 152 and 84 function changes, respectively.
    \item In the \textbf{fourth and fifth rounds}, three researchers analyzed an additional 414 function changes in total.
\end{itemize}

In total, we analyzed 775 function changes, which exceeds the statistically significant sample size (385) with a confidence level of 95\% and a margin of error of 5\%. After three rounds of analysis, all function changes were analyzed by two independent researchers. The Kappa's agreement score was calculated based on the agreement between these two researchers for the entire set of 775 function changes. The obtained Kappa's agreement score of 0.681 indicates substantial agreement among the researchers.

\paragraph{LLM Performance Evaluation}
Following our manual analysis, we proceeded to assess the performance of several leading LLMs in identifying vulnerability-fixing changes. Our study incorporated some of the most widely recognized models, including GPT3.5, GPT4, GPT4o, Claude 3.5 Sonnet, and Gemini 1.5 Pro, for their widespread popularity and cutting-edge performance. It is important to note that the specific versions of the closed-source models employed in our experiments. For GPT3.5, we use the chatgpt-35-0301 version, while for GPT4, we deploy the gpt-4-0314 version. In the case of GPT4o, we adopt the gpt-4o-2024-08-06 version. 

\begin{table}[ht]
\centering
\caption{Comparison of F1-Scores (\%) Across Different Models}
\label{tb:comparison_llm}
\resizebox{0.55\textwidth}{!}{
\begin{tabular}{@{}lcccccc@{}}
\toprule
Model & GPT3.5 & GPT4 & GPT4o & Claude & Gemini \\ 
\midrule
F1-score & 75.85 & \textbf{82.24} & 74.73 & 70.91 & 71.77 \\
\bottomrule
\end{tabular}
}
\end{table}

The results of our evaluation are presented in Table \ref{tb:comparison_llm}, which displays the F1-scores achieved by each model in identifying vulnerability-fixing changes. GPT4 demonstrated superior performance with the highest F1-score of 82.24\%, followed by GPT3.5 at 75.85\% and GPT4o at 74.73\%. Both Claude 3.5 Sonnet and Gemini 1.5 Pro achieved F1-scores of 70.91\% and 71.77\%. These results suggest that while all models show competence in identifying vulnerability-fixing changes, GPT4 maintains a notable edge in this specific task.

\subsection{Ablation Studies}
\label{sec:ablation}

We perform an ablation study by comparing different variants of \tool. \cref{tb:ablation_1} shows the comparison results across different inputs. The largest numbers are highlighted in bold. The variants considered in the study are as follows:

\paragraph{w/o Context}
Our tool without the contextual information of other changed functions in the commit. Without this information, the model may miss important relationships between interrelated changes, impacting the effectiveness of vulnerability detection. The F1-score drops from 82.24\% to 77.64\%, showing a decrease of 4.6\%. These results confirm that contextual information improves detection accuracy.

\begin{table}[ht]
\centering
\caption{Ablation Study Results Comparing F1-Scores (\%) Across Different Inputs}
\label{tb:ablation_1}
\resizebox{0.47\textwidth}{!}{
\begin{tabular}{@{}lc@{}}
\toprule
 & F1-score (\%) \\ 
\midrule
\tool & \textbf{82.24} \\
w/o Context & 77.64 \\
w/o Commit Message & 81.17 \\
w/o Commit Message \& Context & 76.00 \\
\bottomrule
\end{tabular}
}
\end{table}

\paragraph{w/o Commit Message}
Our tool without utilizing commit messages in the analysis. Commit messages often contain valuable information about the nature and purpose of changes. Without them, the F1-score decreases by 1.07\% (from 82.24\% to 81.17\%). While the impact is smaller than removing context, this still demonstrates the value of commit message information in vulnerability detection.

\paragraph{w/o Commit Message \& Context}
Our tool without both commit messages and contextual information. This represents the most basic configuration, relying solely on function changes. The F1-score drops to 76.00\%, a decrease of 6.24\% from the full configuration. This substantial reduction highlights the importance of having both commit messages and contextual information for effective vulnerability detection.

\begin{table}[ht]
\centering
\caption{Ablation Study Results Comparing F1-Scores (\%) Between Our 0-4 Output and Simple Binary Output}
\label{tb:ablation_2}
\resizebox{0.31\textwidth}{!}{
\begin{tabular}{@{}lc@{}}
\toprule
 & F1-score (\%) \\ 
\midrule
\tool & 82.24 \\
w/o 0-4 Output & 74.66 \\
\bottomrule
\end{tabular}
}
\end{table}

\paragraph{w/o 0-4 Output}
We then conducted an additional ablation study to evaluate the impact of our 0-4 scoring output design compared to a simple binary output approach, using our best-performing input combination of function changes plus commit message and context. The results are presented in \cref{tb:ablation_2}. The analysis demonstrates that our proposed 0-4 output design improves the detection performance. When replacing the 0-4 output design with a simple binary output, the F1-score decreases from 82.24\% to 74.66\%, showing a reduction of 7.58\%. This substantial difference highlights the effectiveness of our fine-grained vulnerability scoring approach compared to traditional binary classification.

\begin{table}[ht]
\centering
\caption{Ablation Study Results Comparing F1-Scores (\%) With and Without Heuristic Filtering}
\label{tb:ablation_3}
\resizebox{0.30\textwidth}{!}{
\begin{tabular}{@{}lc@{}}
\toprule
 & F1-score (\%) \\ 
\midrule
\tool & 82.24 \\
w/o Heuristics & 78.97 \\
\bottomrule
\end{tabular}
}
\end{table}

\paragraph{w/o Heuristics} 
\tool without the heuristic approach that filters out test-related changes before processing. Without these heuristics, the model may waste computational resources and potentially be misled by test code modifications that are not actual vulnerability fixes. The F1-score decreases from 82.24\% to 78.97\%, showing a reduction of 3.27\%. This decline demonstrates that our heuristic approach effectively improves the tool's ability to identify genuine vulnerability-fixing changes by focusing the analysis on the most relevant code modifications.

\section{Threats to Validity}
\label{sec:threats}

\paragraph{Internal Validity}
Internal validity refers to the extent to which a study establishes a causal relationship between the independent and dependent variables, free from systematic errors and bias.
One potential threat to internal validity arises from the manual labeling process used to create our evaluation dataset. To mitigate this threat, we employed multiple researchers for the labeling process and calculated Cohen's Kappa agreement score (0.681), indicating substantial inter-rater reliability. Additionally, we conducted multiple rounds of analysis and cross-validation among researchers to ensure consistency.
Another internal threat stems from the probabilistic nature of LLMs, which can lead to performance variations between runs. To address this, we conducted three runs for each experiment and reported the average results.

\paragraph{External Validity}
External validity refers to the extent to which research findings can be generalized to other contexts and settings. To ensure generalizability across programming languages, our dataset includes code from multiple popular languages (Java, Python, C, JavaScript, C\#, C++). Additionally, we selected projects with diverse development practices and team sizes that mirror typical software development environments, enhancing the applicability of our findings to real-world scenarios.

\paragraph{Construct Validity}
Construct validity concerns the extent to which a study's measurements actually represent the intended theoretical constructs. To ensure robust evaluation, we employed a comprehensive set of standard metrics (F1-score, precision, recall, accuracy, and correctness) commonly used in vulnerability detection research.

\section{Related Work}
\label{sec:related_work}

In this section, we review existing research related to vulnerability datasets, methods for improving dataset accuracy, and the application of machine learning models in vulnerability detection.

\subsection{Vulnerability Datasets}

The creation and curation of high-quality vulnerability datasets is critical for advancing automated vulnerability detection techniques. Several notable datasets have been developed in recent years: BigVul \cite{fan2020ac} and CodeXGLUE \cite{zhou2019devign,lu2021codexglue} are large-scale datasets containing over 10K vulnerable functions. However, as noted by Chen et al. \cite{chen2023diversevul}, these datasets suffer from low correctness rates of around 25\%, meaning a significant portion of the labeled vulnerabilities may be inaccurate. More recent efforts have improved dataset quality. CrossVul \cite{nikitopoulos2021crossvul} and CVEFixes \cite{bhandari2021cvefixes} achieved correctness rates of 47.8\% and 51.7\% respectively \cite{ding2024vulnerability}. DiverseVul \cite{chen2023diversevul} further improved on this with a 60\% correctness rate across nearly 19,000 functions. The current state-of-the-art in terms of dataset quality is represented by SVEN \cite{he2023large} and PrimeVul \cite{ding2024vulnerability}, which report very high correctness rates of 94\% and 86\% respectively. However, the SVEN dataset is notably small, containing only 803 functions, primarily due to the constraints of manual analysis. PrimeVul leverages NVD descriptions to match function names, but this method limits its applicability to VFCs that do not correspond to NVD entries. 

Our work aims to bridge the gap between dataset size and quality. \dataset achieves a correctness rate of 90.6\% (comparable to SVEN and PrimeVul) while maintaining a larger scale of 8,198 functions. Importantly, \dataset is derived from GitHub data rather than relying solely on NVD entries, making it a valuable complement to existing high-quality datasets.

\subsection{Automated Vulnerability Detection Techniques}

Research in automated vulnerability detection has seen significant progress, evolving from traditional static analysis techniques to more sophisticated machine learning approaches: Early work focused on static analysis tools that use predefined rules to identify potential vulnerabilities \cite{livshits2005dynamine, jovanovic2006pixy}. While effective for certain types of vulnerabilities, these approaches often suffer from high false positive rates and struggle with complex, context-dependent vulnerabilities. Machine learning techniques have emerged as a promising direction for improving vulnerability detection. Supervised learning approaches have been applied to classify code as vulnerable or benign based on features extracted from source code \cite{scandariato2014predicting}. These methods have shown improved accuracy over traditional static analysis but are highly dependent on the quality of training data. Deep learning models have recently gained traction in this domain. Wang et al. \cite{wang2016automatically} proposed using deep belief networks for vulnerability detection, while Li et al. \cite{li2018vuldeepecker} introduced VulDeePecker, a neural network-based approach for detecting vulnerabilities in source code. More recent work has explored the use of graph neural networks \cite{zhou2019devign} and transformer-based models \cite{chen2023diversevul} for vulnerability detection, showing promising results.

\subsection{Dataset Cleaning and Noise Reduction}

The challenge of noisy labels in security datasets has been recognized in several studies: Ding et al. \cite{ding2024vulnerability} highlighted the issue of noise in vulnerability datasets, reporting that existing datasets often contain 40\% to 75\% noisy data. They proposed a method to improve dataset quality by correlating function names from commit logs with NVD descriptions. Chen et al. \cite{chen2023diversevul} addressed the dataset noise problem by developing a multi-stage filtering process to create a more diverse and accurate vulnerability dataset. 

Our work contributes to this area by proposing the first automatic approach for identifying and filtering out non-vulnerability-related changes in commits, without requiring NVD entry links.
We provide a detailed analysis of the types of changes affecting non-vulnerable functions commonly found in VFCs and demonstrate the effectiveness of \tool in reducing noise in the resulting dataset.

\section{Conclusion and Future Work}
\label{sec:conclusion}

In this paper, we addressed the critical challenge of accurately identifying vulnerability-fixing changes within vulnerability-fixing commits (VFCs), a task essential for improving the effectiveness of machine learning models in automated vulnerability detection. Our comprehensive study revealed that a significant portion of changes within VFCs are not directly related to fixing vulnerabilities, with \emph{Test-Related Changes} and \emph{Bug Fixes} accounting for 41.2\% and 38.2\% of non-vulnerability changes, respectively. This insight informed the development of our LLM heuristic approach - \textbf{\tool}, which demonstrated superior performance in identifying genuine vulnerability fixes in VFCs. Notably, GPT-4, enhanced with our heuristic method, achieved an F1-score of 0.82, making it the first approach to automatically identify genuine vulnerability fixes in VFCs, without requiring NVD entry links.

We created \textbf{\dataset}, a new high-quality vulnerability dataset, by analyzing 5,352,105 commits from 127,063 GitHub repositories. Using \tool, we filtered out noise from vulnerability-fixing commits (VFCs) in this corpus. \tool has a configurable threshold that can be set based on data cleanliness requirements, allowing us to balance dataset size and quality.
With a threshold of 3, \dataset contains 8,198 vulnerable function pairs achieving 90.6\% \emph{Correctness}, while a stricter threshold of 4 yields 6,368 function pairs with 97.3\% \emph{Correctness} in our test sample. These results are comparable to established datasets such as SVEN (94.0\% \emph{Correctness}) and PrimeVul (86.0\% \emph{Correctness}), while overcoming their limitations. Unlike SVEN, which relies on manual analysis and contains only 803 samples, our approach is scalable. Additionally, while PrimeVul requires NVD entry links, our \tool can analyze all VFCs, including those without such links.
Our evaluation of various LLMs fine-tuned on \dataset revealed its superior generalization capabilities across different datasets. Notably, models fine-tuned on \dataset significantly outperformed PrimeVul-trained models when tested on SVEN, achieving 64.87\% accuracy compared to 55.75\%. When evaluated on PrimeVul, our models demonstrated better performance than those trained on SVEN (58.09\% vs 55.20\%). Most remarkably, models trained on \dataset and tested on PrimeVul achieved higher accuracy than models trained and tested on PrimeVul itself (58.09\% vs 57.19\%). These results underscore \dataset's effectiveness and robustness for model fine-tuning, particularly in its ability to capture generalizable vulnerability patterns that transfer well across different contexts.

In the future, we plan to extend our approach in several key directions. First, we aim to develop techniques that can effectively process long-form software artifacts by exploring hierarchical analysis methods that can maintain model attention across extensive codebases. Given our observation that smaller specialized models sometimes outperform larger ones, we plan to investigate architectures specifically optimized for vulnerability detection tasks. We also intend to explore methods for intelligent context integration, combining commit messages with code diffs and project metadata while respecting model input limitations. Finally, we plan to work on semi-automated dataset curation approaches that leverage both LLM capabilities and expert validation to further improve the quality of vulnerability datasets, building upon the success of our current dataset.

\bibliography{bibliography}
\bibliographystyle{ACM-Reference-Format}

\clearpage
\appendix
\section{Appendix}

\begin{figure}[htb]
\begin{lstlisting}[language=Python]
self.function_patterns = {
    # Java test method patterns
    'java': [
        r'@Test\s+.*?(?:public\s+)?void\s+(\w+)\s*\([^\)]*\)',
        r'@Before\s+.*?(?:public\s+)?void\s+(\w+)\s*\([^\)]*\)',
        r'@After\s+.*?(?:public\s+)?void\s+(\w+)\s*\([^\)]*\)',
        r'@BeforeEach\s+.*?(?:public\s+)?void\s+(\w+)\s*\([^\)]*\)',
        r'@AfterEach\s+.*?(?:public\s+)?void\s+(\w+)\s*\([^\)]*\)'
    ],
    # C/C++ test function patterns
    'cpp': [
        r'TEST\s*\(\s*(\w+)\s*,\s*(\w+)\s*\)',
        r'TEST_F\s*\(\s*(\w+)\s*,\s*(\w+)\s*\)',
        r'TEST_P\s*\(\s*(\w+)\s*,\s*(\w+)\s*\)'
    ],
    # C# test method patterns
    'csharp': [
        r'\[Test(?:Case)?\]\s*.*?(?:public\s+)?void\s+(\w+)\s*\([^\)]*\)',
        r'\[TestMethod\]\s*.*?(?:public\s+)?void\s+(\w+)\s*\([^\)]*\)',
        r'\[Fact\]\s*.*?(?:public\s+)?void\s+(\w+)\s*\([^\)]*\)',
        r'\[Theory\]\s*.*?(?:public\s+)?void\s+(\w+)\s*\([^\)]*\)'
    ],
    # JavaScript test function patterns
    'javascript': [
        r'test\s*\(\s*[\'"].*?[\'"]\s*,\s*(?:function|\([^\)]*\)\s*=>)',
        r'it\s*\(\s*[\'"].*?[\'"]\s*,\s*(?:function|\([^\)]*\)\s*=>)',
        r'describe\s*\(\s*[\'"].*?[\'"]\s*,\s*(?:function|\([^\)]*\)\s*=>)',
        r'beforeEach\s*\(\s*(?:function|\([^\)]*\)\s*=>)',
        r'afterEach\s*\(\s*(?:function|\([^\)]*\)\s*=>)'
    ],
    # Python test function patterns
    'python': [
        r'@pytest\.mark\..*?\s*def\s+(\w+)\s*\([^\)]*\):',
        r'@unittest\..*?\s*def\s+(\w+)\s*\([^\)]*\):',
        r'def\s+(test_\w+)\s*\([^\)]*\):',  # unittest style test_* functions
        r'@pytest\.fixture\s*.*?def\s+(\w+)\s*\([^\)]*\):',
        r'@pytest\.(?:mark\.)?parametrize\s*.*?def\s+(\w+)\s*\([^\)]*\):'
    ]
}

# Test indicators specifically for matching file names
self.test_indicators = [
    r'^test',  # File starts with 'test'
    r'test$',  # File ends with 'test'
    r'Test',  # File contains 'Test'
    r'_test$',  # File ends with '_test'
    r'^test_',  # File starts with 'test_'
    r'_Test$',  # File ends with '_Test'
    r'^Test'  # File starts with 'Test'
]
\end{lstlisting}
\caption{Regular expression patterns for identifying test functions and test files across multiple programming languages (Java, C++, C\#, JavaScript, and Python). The patterns capture both function declarations and test file naming conventions commonly used in various testing frameworks.}
\label{fig:heursitcs}
\end{figure}

\begin{figure}[htb]
\begin{lstlisting}[language=Tex]
(*@\textbf{The Prompt Produces a Binary Output:}@*)
As a cybersecurity expert, analyze the provided "Original" and "Revised" code snippets from a commit, along with the commit message and other functions in the same commit. The "Original" code represents the state before the changes, while the "Revised" code represents the state after the changes. (*@\textbf{Determine if the changes are focused on fixing vulnerabilities; if so, output 1, otherwise output 0}@*). The length of the code snippet should not influence your assessment; concentrate on evaluating the logic line by line.

- A score of 0 indicates that the changes made from the "Original" code to the "Revised" code (*@\textbf{do not address vulnerability fixes}@*).
- A score of 1 indicates that the changes made from the "Original" code to the "Revised" code (*@\textbf{are aimed at fixing vulnerabilities}@*).

Commit Message:
{commit}

Original code snippet (code before changes):
{original}

Revised code snippet (code after changes):
{revised}

Here are the other functions in the same commit:
{context}
\end{lstlisting}
\caption{The Prompt Produces a Binary Output, Indicating Whether Changes are Vulnerability-Fixing or Not.}
\label{fig:binary_output}
\end{figure}

\end{document}